\documentclass[conference]{IEEEtran}
\IEEEoverridecommandlockouts
\usepackage{graphicx}
\usepackage{booktabs}
\usepackage{listings}
\usepackage{xcolor}
\usepackage{hyperref}
\usepackage{amsmath}
\usepackage{algorithm}
\usepackage{algpseudocode}
\usepackage{multirow}

\usepackage{array}
\newcolumntype{P}[1]{>{\raggedright\arraybackslash}p{#1}}

\lstdefinestyle{fortran}{
  language=[90]Fortran,
  basicstyle=\ttfamily\footnotesize,
  keywordstyle=\color{blue}\bfseries,
  commentstyle=\color{gray}\itshape,
  stringstyle=\color{red},
  numbers=left,
  numberstyle=\tiny\color{gray},
  stepnumber=1,
  numbersep=5pt,
  frame=single,
  breaklines=true,
  captionpos=b,
  morekeywords={acc,kernels,loop,independent,reduction,routine,seq,wait},
}

\lstdefinestyle{bash}{
  language=bash,
  basicstyle=\ttfamily\footnotesize,
  frame=single,
  breaklines=true,
}
\begin{document}

\title{Validation-Centric AI-Assisted GPU Porting of a 250,000+ Line Legacy Weather Simulation Code}

\author{
\IEEEauthorblockN{Tetsuya Hoshino}
\IEEEauthorblockA{
\textit{Information Technology Center}\\
\textit{Nagoya University}\\
Aichi, Japan\\
hoshino@cc.nagoya-u.ac.jp}
\and
\IEEEauthorblockN{Masaya Kato}
\IEEEauthorblockA{
\textit{Institute for Space-Earth}\\
\textit{Environmental Research}\\
\textit{Nagoya University}\\
Aichi, Japan\\
}
\and
\IEEEauthorblockN{Kazuhisa Tsuboki}
\IEEEauthorblockA{
\textit{Institute for Space-Earth}\\
\textit{Environmental Research}\\
\textit{Nagoya University}\\
Aichi, Japan\\
} 
\and
\IEEEauthorblockN{Daichi Mukunoki}
\IEEEauthorblockA{
\textit{Information Technology Center}\\
\textit{Nagoya University}\\
Aichi, Japan\\
}
\and
\IEEEauthorblockN{Takahiro Katagiri}
\IEEEauthorblockA{
\textit{Information Technology Center}\\
\textit{Nagoya University}\\
Aichi, Japan\\
}
\and
\IEEEauthorblockN{Toshihiro Hanawa}
\IEEEauthorblockA{
\textit{Information Technology Center}\\
\textit{The University of Tokyo}\\
Chiba, Japan\\
}
}

\maketitle

\begingroup
\footnotesize
\noindent\textit{This work has been submitted to the IEEE for possible publication.
Copyright may be transferred without notice, after which this version may no longer be accessible.}
\par
\endgroup
\vspace{1mm}

\begin{abstract}
Recent advances in large language models have made CLI-based AI agents a practical tool for accelerating GPU porting of large legacy scientific applications. Such applications, however, are not merely old code bases; they are scientific assets whose credibility has been accumulated through long-term development, comparison with observations, and use in domain studies. GPU porting must therefore preserve this scientific validity while adapting the implementation to GPU-centric HPC systems.
This paper presents a validation-centric AI-assisted GPU porting workflow through a case study of CReSS, a legacy Fortran weather simulation code with more than 250,000 lines. The workflow uses an AI agent to extract OpenMP regions, generate dump-based kernel benchmarks from physically meaningful simulation states, apply OpenACC transformations, and validate results through element-wise comparison with dumped reference data and application-level validation.
Using a real typhoon simulation, the workflow produced numerically validated GPU implementations for 162 target kernels and achieved a 5.1$\times$ application-level speedup within practical wall-clock development cost.
In particular, it detected numerical discrepancies in five kernels caused by floating-point and intrinsic-function differences, including threshold-sensitive branch divergence and cancellation effects, enabling feedback to the application developers. 
The case study suggests that, for large legacy scientific applications requiring dump-based validation, practical AI-assisted GPU porting must manage session-spanning context, runtime-state reconstruction, and costly recovery from small static-analysis omissions. These findings demonstrate that AI-assisted GPU porting requires not only code generation, but validation-centric workflow design.

\end{abstract}

\begin{IEEEkeywords}
GPU porting, large language models, code generation, OpenACC, Fortran,
weather simulation
\end{IEEEkeywords}

\section{Introduction}

Recent advances in large language models (LLMs) and AI agents have made it increasingly feasible to support software development tasks such as code analysis, transformation, testing, and debugging. 
In the HPC community, recent studies have reported promising results for LLM-assisted code generation, directive insertion, and parallel-code translation on kernels or limited workflow units~\cite{godoy_llm_hpc,hpccoder,unipar,geofem_llm}.
At the same time, HPC systems are rapidly moving toward GPU-centric architectures, making the GPU porting of large scientific applications developed for CPU-based systems an important challenge. 
Such applications often consist of hundreds of thousands of lines of Fortran or C/C++ code, hybrid MPI/OpenMP parallelization, complex physical models, and long maintenance histories. 
Manually porting them to GPUs requires substantial expertise and development effort, motivating the use of AI agents to reduce GPU-porting cost.

However, legacy scientific applications are not merely old code bases. 
In many cases, they are computational models whose scientific validity has been accumulated through long-term development, comparison with observations, and repeated use in domain studies.
For example, CReSS~\cite{Tsuboki2002,Tsuboki2008,Tsuboki2023}, the cloud-resolving weather simulation model targeted in this study, has been developed since 1998 and has been used in studies of tropical cyclones, heavy rainfall, convective systems, and other severe weather phenomena.
Its reliability has been improved through comparison with observations and continuous model refinement.
For such applications, the goal of GPU porting is not to regenerate a new code base from scratch using AI. 
Rather, the goal is to adapt the existing implementation to GPU-centric HPC systems while preserving the scientifically validated behavior established over many years.

This requirement makes verification and validation central to AI-assisted GPU porting. 
Code generated or modified by an AI agent must be checked against the expected behavior of the existing application. 
In addition, numerical differences introduced by GPU execution must be assessed to determine whether they are acceptable variations or implementation defects. 
Since meaningful kernel inputs in scientific applications are produced through initialization, physical processes, and configuration-dependent execution paths, kernel-level validation also requires reconstructing runtime states obtained from actual simulations.

A validation-centric approach is not new to GPU porting itself. 
A conservative way to port a large scientific application is to isolate target kernels, construct benchmarks whose outputs can be compared with CPU references, validate each GPU kernel before integration, and then validate the integrated application. 
This approach localizes errors and helps developers distinguish numerical variation from implementation defects. 
However, applying it exhaustively to a large application is extremely expensive when performed manually, because it requires repeated variable extraction, runtime-state acquisition, benchmark generation, output comparison, and integration testing across many kernels. 
In this work, we use the AI agent not to bypass this validation process, but to reduce the wall-clock development time required to execute such a validation-centric workflow.
Our goal is to obtain a numerically validated GPU implementation within practical wall-clock cost while preserving the scientific validity of the existing application.

We present a validation-centric AI-assisted GPU porting workflow for CReSS, a 250,000+ line legacy Fortran weather simulation code. 
The workflow extracts OpenMP regions, constructs dump-based kernel benchmarks from physically meaningful simulation states, applies OpenACC transformations, and validates the results at both kernel and application levels.

This paper makes three contributions. First, it presents a validation-centric AI-assisted GPU porting workflow for CReSS, aimed at reducing the wall-clock development time to a numerically validated GPU implementation while preserving scientific validity. 
Second, it demonstrates the workflow on 162 kernels in a real typhoon scenario, satisfying application-level validation criteria, achieving a 5.1$\times$ speedup, and detecting five numerical discrepancies for developer feedback. 
Third, we analyze workflow cost and agent-dependent failure modes observed in the CReSS case. The analysis shows that runtime-state reconstruction, session-spanning context management, and cost-aware recovery are central to stabilizing the workflow and reducing wall-clock development cost.

\section{Verification and Validation Requirements in AI-Assisted GPU Porting}
\label{sec:verification_requirements}

This section summarizes the verification and validation requirements that arise in AI-assisted GPU porting of large scientific applications. 
In this paper, verification refers to checking whether code generated or modified by an AI agent preserves the expected behavior of the existing application, whereas validation refers to assessing whether numerical differences introduced by GPU execution are acceptable for the target scientific simulation.
For applications such as CReSS, executing the full application as a black box is insufficient for localizing AI-generated changes and interpreting numerical differences.
The workflow must localize changes, compare results against the existing CPU implementation, and reconstruct meaningful runtime states for kernel-level validation.

\subsection{Verification of AI-Generated Transformations}
\label{subsec:ai_generated_verification}

Code generated or modified by an AI agent must be checked against the expected behavior of the existing application.
Successful compilation or syntactically plausible code does not guarantee correctness for a scientific application. 
Large applications often contain assumptions that are not visible from a local code fragment, such as calling conditions, array shapes, physics options, build options, and data states established by earlier execution phases.

In this work, we do not evaluate the autonomy of the AI agent itself. 
Instead, we focus on a workflow in which AI-generated modifications are made verifiable. 
Application-level output alone is often insufficient: if the final result changes, it can be difficult to determine which transformation caused the difference and whether the difference is a numerical effect or an implementation defect. 
Therefore, the porting task should be decomposed into diagnostic units whose outputs can be compared with reference results from the existing CPU implementation.

This decomposition is not based on the assumption that AI agents cannot be autonomous. Rather, it is a design choice for large scientific applications: the verification scope must be localized, the results must remain interpretable to human developers, and each AI-agent session should handle a bounded amount of code, state, and validation output. 
This locality reduces context-window pressure and makes recovery across session boundaries easier.
However, once a kernel is extracted, it must be executed under a context consistent with the original application. 
This requirement leads to runtime-state reconstruction, discussed in Section~\ref{subsec:runtime_state_reconstruction}.

\subsection{Numerical Validation after GPU Porting}
\label{subsec:numerical_validation}

GPU porting also requires validation of numerical results.
Floating-point differences are not unique to GPUs; they are inherent to parallel numerical computing. 
However, GPU porting often changes the exposed parallelism. 
Inner summations or local reduction-like computations that were sequential within each CPU thread may be parallelized across GPU threads. 
As a result, execution order, rounding behavior, and intrinsic-function implementations may differ between the CPU and GPU versions.

Therefore, bitwise disagreement between CPU and GPU outputs does not immediately imply an implementation error.
At the same time, such differences cannot simply be ignored.
Developers must determine whether the observed differences are acceptable numerical variations or defects introduced by GPU porting. 
This distinction is particularly important for time-dependent simulations such as weather models, where small numerical differences may accumulate over time and eventually affect conditional branches or physical processes.

For this reason, both kernel-level validation and application-level validation are needed. 
Kernel-level validation enables fine-grained diagnosis of where numerical differences arise, including differences caused by intrinsic functions or changes in evaluation order. 
Application-level validation is also necessary because a difference that appears acceptable for a single snapshot may still affect the long-term behavior of the integrated simulation.

\subsection{Runtime-State Reconstruction for Kernel-Level Validation}
\label{subsec:runtime_state_reconstruction}

Kernel-level validation requires input data and reference output data for each target kernel.
However, in large scientific applications, such inputs cannot always be constructed artificially in a meaningful way. 
A kernel input state is often formed through initialization, time-dependent physical processes, configuration-dependent execution paths, memory allocation, and binary-data conventions used for dumped data.
These conditions are part of the execution context of the original application.

Consequently, tests based only on random inputs or small synthetic data may fail to reproduce numerical states or control-flow conditions that occur in actual simulations. 
This issue is especially important for weather simulations, where physical processes evolve over time and interact with each other. 
Validation using only an initial timestep or artificial input may miss behavior that appears only after the simulation state has physically evolved.

In this work, we therefore use runtime states obtained from actual simulation execution to construct kernel benchmarks.
The CPU version of the application is executed, and the input and reference output of target kernels are dumped. 
The dump data are then used to build independent kernel benchmarks, and the GPU output is compared element by element against the reference output.

Runtime-state reconstruction is itself a major challenge. 
It is not sufficient to enumerate variables referenced by a kernel.
The workflow must also determine under which conditions each variable is valid, how arrays are allocated, and which binary-data conventions, such as byte-order conversion, are required to interpret dumped data correctly.
Such information may depend on initialization routines, physics options, conditional allocation, and global configuration rather than on the syntax of the target loop alone.

These requirements motivate the CReSS workflow described in the next section.
\section{Case Study Design: A Validation-Centric Workflow for CReSS}
\label{sec:case-study}

This section describes how the requirements in Section~\ref{sec:verification_requirements} are instantiated in our CReSS case study. 
The workflow is designed to port the existing CReSS implementation to GPUs while preserving the behavior of the original CPU version. 
It uses two levels of validation: kernel-level validation using dumped reference data from the original simulation, and application-level validation after integration. 
The AI agent assists repetitive artifact generation, while human developers define validation requirements, manage specifications, and interpret numerical differences.

\subsection{Target Application and Validation Scenario}
\label{subsec:cress}

The target application is CReSS (Cloud Resolving Storm Simulator), a legacy weather simulation code written in Fortran with hybrid MPI/OpenMP parallelization. 
The code base consists of 599 Fortran 90 source files, approximately 260,000 lines of code, and 387 OpenMP parallel regions.
These OpenMP regions form the primary computational units considered in this study.

Many computational regions in CReSS are already expressed as OpenMP-parallel loops, and the dominant kernels are primarily memory-bandwidth bound. 
This makes directive-based GPU porting using OpenACC a practical first target. 
At the same time, the application contains a large number of kernels that must be validated and integrated together, making the workflow substantially more difficult than the transformation of a single kernel.

We use a real typhoon simulation scenario over the western Pacific in September 2022. 
The simulation uses a grid of $899 \times 899 \times 128$, corresponding to approximately 100 million grid points, with a horizontal resolution of approximately 2 km. 
Major physical processes in CReSS, including cloud microphysics, radiation, turbulence, and surface processes, are enabled. 
Among the 387 OpenMP parallel regions, 162 kernels are executed in this scenario and are used as GPU-porting targets.

The horizontal resolution and domain size are representative of high-resolution regional typhoon simulations with CReSS.
The validation simulation is shortened to 30 minutes of simulated time, or 360 timesteps, to make repeated validation feasible while still exercising the target physical processes and allowing numerical differences to accumulate. 
Initial and boundary conditions are derived from real Grid Point Value (GPV) data.

For application-level validation, we use the maximum and minimum pressure perturbation values. 
Let \(p_{\max}\) and \(p_{\min}\) denote the CReSS variables \texttt{ppmax} and \texttt{ppmin}, respectively. 
For the target typhoon simulation scenario, the CReSS developers provided reference values \(p_{\max}^{\rm ref}=1.140663\times10^3\) and \(p_{\min}^{\rm ref}=-3.927225\times10^3\). 
We define
\[
a_1 =
\frac{|p_{\max} - p_{\max}^{\rm ref}|}
     {|p_{\max}^{\rm ref}|},
\qquad
a_2 =
\frac{|p_{\min} - p_{\min}^{\rm ref}|}
     {|p_{\min}^{\rm ref}|}.
\]
The integrated GPU application is considered valid when both \(a_1\) and \(a_2\) are below \(10^{-4}\).

\subsection{Overview of the Validation-Centric Workflow}
\label{subsec:workflow-overview}

\begin{table*}[t]
\caption{Input and output artifacts in the validation-centric AI-assisted GPU porting workflow.}
\label{tab:workflow_artifacts}
\centering
\footnotesize
\setlength{\tabcolsep}{3pt}
\renewcommand{\arraystretch}{1.08}
\begin{tabular}{
  P{0.18\textwidth}
  P{0.38\textwidth}
  P{0.38\textwidth}
}
\hline
\textbf{Phase} &
\textbf{Input artifacts} &
\textbf{Output artifacts} \\
\hline

Code meta-review &
Original CReSS source code and OpenMP parallel regions &
Annotated OpenMP regions and potential GPU-porting barriers \\

Profiling &
Annotated source code and CPU validation scenario &
Execution profile, invocation counts, and active target kernel list \\

Kernel extraction and CPU benchmark generation &
Annotated source code and active target kernel list and CPU validation scenario &
Dump-instrumented CPU application and dumped input/reference-output states and verified standalone CPU kernel benchmarks \\

Kernel-level GPU transformation &
Dumped input/reference-output states and verified CPU kernel benchmarks &
Validated OpenACC GPU benchmark kernels and numerical discrepancy records \\

Integration &
Validated GPU kernels and original CReSS source code &
Validated integrated GPU application with conditional compilation \\

Performance validation &
Validated integrated GPU application and GPU benchmarks &
Nsight profile, Unified Memory behavior report, and revalidated kernel-local revisions \\
\hline
\end{tabular}
\end{table*}

The workflow consists of six phases: code meta-review, profiling, kernel extraction and CPU benchmark generation, kernel-level GPU transformation, integration, and performance validation. 

Table~\ref{tab:workflow_artifacts} summarizes the artifacts passed between phases. 
The early phases identify target kernels and collect structural information, while formal validation is performed after CPU benchmark generation, GPU transformation, and integration. 
CPU benchmarks are verified against dumped reference data, GPU benchmarks are validated against the same reference data, and the integrated application is validated using the metrics in Section~\ref{subsec:cress}.
Workflow rules, compiler options, binary-I/O conventions, and recovery policies are externalized into specifications so that validation requirements can be preserved across interrupted AI-agent sessions.

\subsection{Code Meta-Review and Target Scoping}
\label{subsec:meta-review}

Before benchmark generation, we perform a code meta-review to make the structure of many OpenMP regions visible to the human developer and to identify structural obstacles to directive-based GPU porting. 

For each OpenMP parallel region, the agent is instructed to inspect features visible from the region itself, such as uses of \texttt{omp\_get\_thread\_num}, OpenMP synchronization constructs such as \texttt{atomic} or \texttt{critical}, function or subroutine calls inside the region, and writes to global or module variables. 
These features are recorded as annotations in the source code. 
The resulting annotations serve as a risk inventory for target selection and later specification design.
They help the human developer confirm whether the executed OpenMP regions are suitable for OpenACC-based GPU porting and whether any region requires special treatment before benchmark generation.

For the executed regions in the target validation scenario, the meta-review did not identify a structural obstacle that prevented proceeding to benchmark generation.

\subsection{Runtime-State Reconstruction and Kernel Benchmarks}
\label{subsec:runtime-state}
A central component of the workflow is the construction of kernel benchmarks from runtime states of the original CPU simulation. 
Using the active target kernel list and invocation counts obtained from profiling, the workflow determines the dump point for each target OpenMP region. 
For each target region, the AI agent extracts the variables required by the region and generates a standalone benchmark program. 
The input data and reference output data for the benchmark are obtained by executing the original CPU simulation and dumping the program state at the selected dump point.

This dump-based approach is used because synthetic inputs are insufficient for meaningful validation in CReSS. 
Kernel inputs are produced through long initialization procedures and time-dependent physical processes, rather than being isolated artificial values. 
Therefore, each kernel must be validated under a runtime context consistent with the original simulation.
In this study, the selected dump point is the last invocation of each target kernel during the 360-step validation simulation, rather than the first invocation or an artificial input state.
This allows validation to use physically evolved states in which major physical processes and accumulated numerical effects are more likely to appear.

For each extracted kernel, we use the same element-wise error metric for two purposes: CPU benchmark verification and GPU benchmark validation. 
First, the generated CPU benchmark is verified by comparing its output with the dumped CPU reference output. 
Second, after OpenACC transformation, the GPU benchmark is validated against the same dumped reference output. 
Because the target CReSS configuration uses single precision, we set the tolerance to \(\tau=10^{-5}\) in both stages. 
The interpretation of a failure differs between the two stages: for the CPU benchmark, exceeding the tolerance is treated as a failure in kernel extraction, runtime-state reconstruction, or dump replay; for the GPU benchmark, it triggers local diagnosis of the OpenACC transformation and, if no implementation error is found, human interpretation as a numerical discrepancy.

Let $z_{i,j,k}$ be the output under this check, and let $\hat{y}_{i,j,k}$ be the dumped CPU reference value. 
We define
\begin{equation}
e_{i,j,k} =
\frac{\left| z_{i,j,k} - \hat{y}_{i,j,k} \right|}
     {\max\!\left( \left| \hat{y}_{i,j,k} \right|, \epsilon_{\min} \right)},
\label{eq:elemerr}
\end{equation}
where $\epsilon_{\min} = 10^{-20}$ bounds the denominator from below so that the relative error remains computable when the reference value is at or near zero. 
The output passes the element-wise check when $\max_{i,j,k} e_{i,j,k} \le \tau$, where $\tau = 10^{-5}$.

The workflow uses a \texttt{variable\_list} as an intermediate artifact for runtime-state reconstruction. 
The \texttt{variable\_list} records variable names, type and array information, access roles in the target region, and data-validity conditions required for dump and replay. 
The detailed failure modes caused by incomplete \texttt{variable\_list} entries are discussed in Section~\ref{subsec:runtime_state_failures}.

\subsection{GPU Transformation and Integration}
\label{subsec:gpu-integration}

The verified CPU benchmark is retained as a reference artifact, and the AI agent derives a separate OpenACC GPU benchmark from its copy. 
In this workflow, we instruct the agent to use an OpenACC \texttt{kernels} region and to expose candidate loop-level parallelism using \texttt{loop independent} annotations.

This policy is intentionally optimistic but verification-driven within the validation-centric workflow.
Many target regions in CReSS are structured-grid loop nests over three-dimensional fields, and the existing OpenMP parallelization provides evidence of coarse-grained parallelism.
However, we do not assume that all loop levels are independent: some kernels may contain vertical dependencies, sequential accumulations, boundary updates, or state-dependent execution. 
Therefore, a \texttt{loop independent} annotation is treated as a candidate transformation rather than a proof of correctness. 
The transformed kernel is checked immediately using the dump-based kernel benchmark. 
If an incorrect independence assumption changes the output, the difference is detected by element-wise comparison before the kernel is integrated into the full application. 
In such cases, the transformation can be revised, for example by marking the corresponding loop as sequential.

For data management, we use Unified Memory (\texttt{-gpu=managed}) rather than explicit OpenACC data directives. 
This design simplifies the validation workflow by avoiding explicit management of inter-kernel data movement during the initial porting stage. 
The tradeoff is that the resulting implementation may leave optimization opportunities in explicit data management, migration control, and asynchronous execution.

Validated GPU kernels are then integrated into the original application using conditional compilation. 
We use \texttt{\#ifdef} directives to switch between CPU and GPU versions on a per-kernel, or per-kernel-group, basis. 
This enables bisection-style debugging: when application-level validation fails, subsets of GPU kernels can be selectively enabled to narrow down the source of the discrepancy. 
After integration, application-level validation is performed using the pressure-based metrics described in Section~\ref{subsec:cress}.

\subsection{Performance Validation and Kernel-Local Revision}
\label{subsec:performance-validation}

After application-level validation, we perform profiling not as an aggressive optimization phase, but as a performance validation step. 
Since the implementation uses Unified Memory, a numerically correct execution may still exhibit undesirable behavior, such as unexpected CPU--GPU page migration or ineffective GPU parallelization.
Such issues cannot be detected by element-wise numerical validation alone.

We therefore profile the integrated GPU application using NVIDIA Nsight. 
The profiling examines kernel execution time, memory-bandwidth utilization, and Unified Memory data movement. 
When a kernel shows clearly anomalous behavior, we revise only the corresponding GPU benchmark kernel while preserving its input/output interface. 
The revised kernel is then revalidated using the same dumped reference data before being reintegrated into the full application.

This phase is intentionally restricted to kernel-local revisions.
We do not apply aggressive optimizations that introduce inter-kernel dependencies, such as explicit data-region management, \texttt{async} execution, kernel fusion, data-structure changes, or communication overlap. 
These optimizations require a different validation strategy and are therefore treated as future work.

\section{Evaluation of the Ported Application}
\label{sec:evaluation}

This section evaluates the GPU implementation obtained by the validation-centric AI-assisted porting workflow. The purpose of this section is not to evaluate the autonomy of the AI agent, but to assess whether the workflow produced a numerically validated GPU implementation with reasonable performance for the target CReSS validation scenario.

\subsection{Experimental Environment}
\label{subsec:env}

All experiments were conducted on the Miyabi supercomputer system operated by the Joint Center for Advanced High Performance Computing (JCAHPC). The GPU nodes (Miyabi-G) are equipped with NVIDIA GH200 Grace-Hopper Superchips. Each node consists of a 72-core Grace CPU and an NVIDIA H100 GPU connected by NVLink-C2C. The CPU has 120 GB of memory with approximately 512 GB/s memory bandwidth, while the GPU provides 96 GB of HBM memory with a peak bandwidth of approximately 4,022 GB/s.
All validation and performance experiments in this paper use a single Miyabi-G node, consisting of one Grace CPU and one H100 GPU; multi-node execution is outside the scope of this study.

The software environment includes NVIDIA HPC SDK 25.9.
The GPU version is compiled using \texttt{nvfortran} with OpenACC directives and Unified Memory (\texttt{-gpu=managed}). 
The CPU baseline uses the original OpenMP implementation executed on the Grace CPU with 72 threads.
Unless otherwise noted, all experiments use the single-precision CReSS configuration and the target typhoon validation scenario described in Section~\ref{subsec:cress}.

\subsection{Validation Coverage}
\label{subsec:validation-coverage}

Among the 387 OpenMP parallel regions in CReSS, 162 regions are executed in the target validation scenario. 
These 162 regions are the GPU-porting targets in this case study.
For each target region, we first generated a standalone CPU benchmark and verified it against the dumped reference output from the original CPU simulation. 
We then derived an OpenACC GPU benchmark from the verified CPU benchmark and compared the GPU output against the same dumped reference data.

The kernel-level validation is not a purely automatic pass/fail test. If all output elements are within the predefined tolerance, the GPU kernel is accepted directly. 
If some elements exceed the tolerance, the case first enters a local diagnosis and revision loop: the agent and developer check for implementation errors, incorrect OpenACC transformations, or invalid loop-independence assumptions, revise the benchmark if necessary, and rerun validation. 
Only discrepancies that remain after such checks are treated as numerical discrepancies requiring human interpretation, i.e., to determine whether they are acceptable numerical variations rather than implementation defects.

After kernel-level validation, the validated GPU kernels were integrated into the original CReSS code using conditional compilation. 
The integrated application completed the 360-step validation simulation and satisfied the application-level validation criteria based on the application-level metrics defined in Section~\ref{subsec:cress}.

\subsection{Numerical Discrepancies Detected by Kernel-Level Validation}
\label{subsec:numerical-discrepancies}

A key benefit of the workflow is that it exposes numerical discrepancies at the kernel level before they are hidden inside the full application execution. 
In this case study, most GPU kernels were validated without unresolved discrepancies.
However, five kernels produced single-element discrepancies that exceeded the kernel-level tolerance and required human inspection. 
Table~\ref{tab:numerical_discrepancies} summarizes these cases.

\begin{table}[t]
\caption{Numerical discrepancies detected by kernel-level validation.}
\label{tab:numerical_discrepancies}
\centering
\footnotesize
\setlength{\tabcolsep}{3pt}
\renewcommand{\arraystretch}{1.10}
\begin{tabular}{p{0.23\linewidth}p{0.32\linewidth}p{0.35\linewidth}}
\hline
\textbf{Kernel} &
\textbf{Observed discrepancy} &
\textbf{Interpretation} \\
\hline

\texttt{bruntv.f90} &
Threshold-sensitive branch difference in \texttt{t <= tlow} &
Small CPU/GPU difference changed a branch-condition decision \\

\texttt{disptke.f90} &
Difference involving \texttt{exp()}, \texttt{log()}, and \texttt{sqrt()} &
Intrinsic-function difference amplified by cancellation \\

\texttt{cloudcov.f90} &
Large relative error near zero &
Reference value was exactly zero; absolute difference $1.9\times10^{-8}$ \\

\texttt{siadjst.f90} &
Threshold-sensitive branch difference in \texttt{qi > dqi} &
Small numerical difference changed a microphysics adjustment condition \\

\texttt{swadjst.f90} &
Threshold-sensitive branch difference in \texttt{qc > dqc} &
Small numerical difference changed a microphysics adjustment condition \\
\hline
\end{tabular}
\end{table}

These discrepancies were not identified as GPU implementation bugs. 
Instead, they were caused by small numerical differences between CPU and GPU execution that were amplified by threshold-sensitive branches, cancellation, or near-zero relative-error effects. 
For example, in \texttt{bruntv.f90}, the temperature is computed using \texttt{exp()} and \texttt{log()} and then compared with a low-temperature threshold \texttt{tlow}=233.16 K. 
The CPU computation produced \(t=233.16002\) K, making the condition \texttt{t <= tlow} false. 
The GPU computation produced \(t=233.16000\) K (a difference of one ulp in single precision), making the condition true and applying a latent-heat correction term only in the GPU execution.
In \texttt{disptke.f90}, by contrast, intermediate values differed by about 25 ulps, suggesting an intrinsic-function implementation difference rather than a simple one-ulp rounding effect.

The important point is not simply that such differences exist. 
Rather, the kernel-level validation workflow made it possible to identify the responsible kernel, the affected condition, and the numerical mechanism that amplified the small CPU/GPU difference. 
We discussed these cases with the CReSS developers and confirmed that they were acceptable numerical differences for the target validation scenario. 
This feedback would have been much harder to obtain from application-level validation alone.

\subsection{Application-Level Validation}
\label{subsec:application_validation}

Kernel-level validation does not by itself guarantee that the integrated application preserves the expected behavior of the original simulation. 
Therefore, after integrating the validated GPU kernels, we executed the full 360-step validation scenario and evaluated the application-level metrics defined in Section~\ref{subsec:cress}. 
The integrated GPU application satisfied the validation criteria ($a_1 = 1.0 \times 10^{-5}$, $a_2 = 5.6 \times 10^{-5}$).

During integration, the conditional-compilation structure described in Section~\ref{subsec:gpu-integration} allowed us to selectively enable subsets of GPU kernels and localize the kernel group or integration point responsible for application-level validation failures. 
This was necessary because the snapshot-based kernel validation does not cover all execution states that appear during the full time evolution of the simulation. 
We discuss this coverage limitation and the observed integration failures in Section~\ref{subsec:snapshot_limitations}.

\subsection{Performance Validation Results}
\label{subsec:performance-validation-results}

We evaluate the performance of the validated GPU application after the performance-validation and kernel-local revision step described in Section~\ref{subsec:performance-validation}. 
The purpose of this evaluation is not to demonstrate fully optimized GPU performance, but to confirm that the generated OpenACC implementation behaves as a reasonable GPU implementation without obvious performance pathologies.

\begin{figure}[t]
\centering
\includegraphics[width=0.48\textwidth, trim=0 10 0 10, clip]{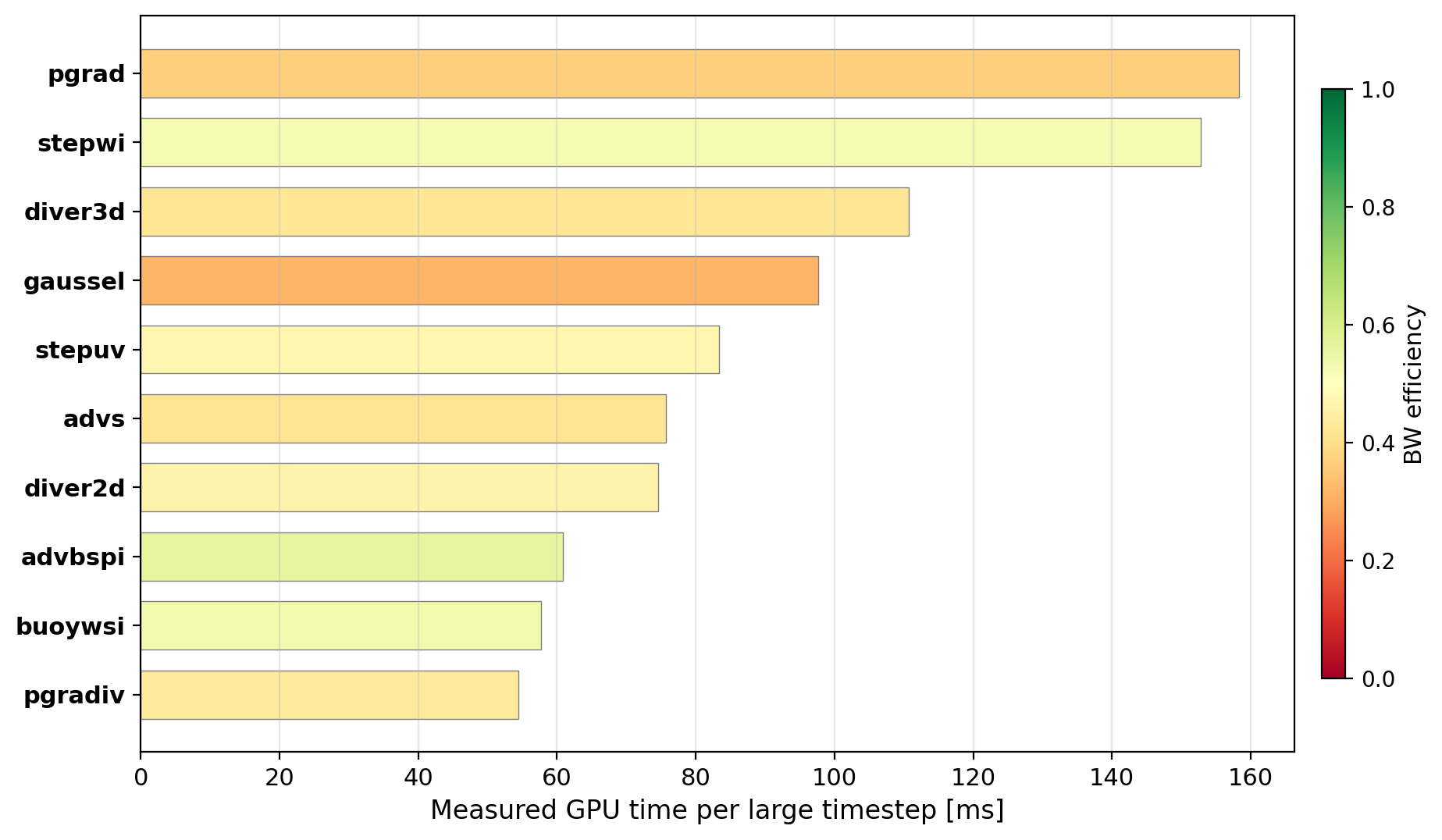}
\caption{Top 10 GPU kernels by measured execution time, with colors indicating memory-bandwidth efficiency relative to the GPU peak.}
\label{fig:gpu_perf}
\end{figure}

We profiled the integrated GPU application using NVIDIA Nsight and inspected all 162 GPU kernels. 
In addition to kernel execution time and memory-bandwidth utilization, we checked Unified Memory migration behavior to detect obvious problems such as unintended CPU execution, excessive CPU--GPU data migration, or missed offloading. 
No such dominant pathology was observed in the target validation scenario.
Across the 162 kernels, memory-bandwidth utilization was roughly 35--60\% of the GPU peak bandwidth, which is reasonable for directive-based OpenACC implementations under this workflow.
Figure~\ref{fig:gpu_perf} shows the top 10 kernels ranked by execution time to visualize the dominant performance components; the profiling and bandwidth assessment were performed for all 162 kernels.

The median execution time per timestep was 9.51~s on the Grace CPU using 72 OpenMP threads and 1.88~s on the GPU, corresponding to a speedup of approximately 5.1$\times$.
Although this is below the roughly 8$\times$ peak memory-bandwidth ratio between the H100 HBM and the
Grace CPU memory, it is reasonable for an OpenACC implementation using Unified Memory under a validation-centric workflow that avoids aggressive inter-kernel optimizations.

\section{Workflow Cost and Agent-Dependent Failure Modes}
\label{sec:workflow_observations}

Section~\ref{sec:evaluation} showed that the proposed workflow produced a numerically validated GPU implementation of CReSS with reasonable performance. 
This section analyzes the cost structure and failure modes observed when executing the validation-centric workflow with Claude Code Opus 4.5--4.6.

The observations should not be interpreted as model-independent limitations of AI agents: the accuracy of static analysis, local recovery, and context management may change with future models or agent architectures. 
Instead, we focus on how agent-dependent errors interact with HPC-specific costs, such as many dump insertion points, large dump data, simulation re-execution, and session boundaries imposed by interactive-job limits.

\subsection{Cost Structure and Reduced Phase-3 Setting}
\label{subsec:cost_structure}

We use the wall-clock time to validation success as the primary practical cost metric. In this case study, the AI agent was executed through interactive jobs on the target HPC system, and the total interactive GPU-node time was approximately 100 hours over about three months, including exploratory iterations, failed attempts, and recovery work.
Because this was an exploratory case study, these interactive sessions were monitored by a human developer for validation and decision-making. 
We therefore regard this node time as an approximate scale of human-supervised development time, not as unattended production compute time.
We do not use this as a quantitative baseline, but it indicates the practical scale of a porting task that would otherwise have been planned on a months-to-years timescale in our practice.

The dominant cost was concentrated in Phase 3: kernel extraction and CPU benchmark generation. 
This phase accounted for more than half of the development wall-clock time. 
Once a standalone kernel benchmark was correctly constructed, the local OpenACC transformation and test cycle was relatively cheap. 
The difficult part was turning the large application into locally testable units without causing expensive simulation re-execution.

Because one dump acquisition requires about one hour of simulation re-execution in our environment, Phase 3 had to be organized as a batched procedure rather than as a per-kernel trial-and-error loop.
Ideally, the workflow first processes all target OpenMP regions: it generates benchmark skeletons, constructs the per-region \texttt{variable\_list}, and inserts dump instrumentation for all regions.
The dump-instrumented CPU validation simulation is then executed once to produce dump data, and the generated CPU benchmarks are verified afterward.

However, if the \texttt{variable\_list} or dump instrumentation is incomplete, the dump run can fail or produce unusable data.
The workflow then enters a costly repair cycle: diagnose the missing variable or data-validity condition, such as physics-option guards that determine whether a full-domain array or a dummy object is valid, repair the \texttt{variable\_list} or dump instrumentation, and rerun the simulation.
In our observations, the agent often repaired only the locally exposed symptom and immediately retried the expensive dump execution, even when similar latent problems remained elsewhere.
The elapsed wall-clock cost of repeated failures did not by itself shift the agent from local repair to broader batch checking.
Thus, the key cost problem in Phase 3 is that local failures in runtime-state reconstruction can be amplified into repeated simulation re-execution unless cost-aware recovery is made explicit.

To evaluate workflow-control strategies under this cost structure, we use a reduced Phase-3 setting. 
The goal of this controlled experiment is to examine how prompt-based and specification-based workflows behave under session boundaries and whether recovery policies can reduce repeated dump execution. 
We fix the model to Claude Code Opus 4.6 in these controlled experiments to remove variability due to model versions.
The reduced setting shortens the simulation from 360 timesteps to 36 timesteps and uses 15 kernels: the top 12 kernels by execution time and three additional kernels that require application-specific conditional handling, discussed in Section~\ref{subsec:runtime_state_failures}. 
Across these 15 kernels, the generated \texttt{variable\_list} files contain 690 entries, including 134 condition-dependent entries, and the resulting dump data amount to approximately 89~GB. 
This setting preserves the essential cost structure of Phase 3---variable extraction, dump acquisition, benchmark generation, and verification---while making repeated agentic workflow experiments feasible.

The use of interactive HPC sessions is not incidental.
Dump acquisition and validation must be performed on the target HPC system because the simulation requires large-scale resources and because the generated GPU code must be tested in the target execution environment. 
On Miyabi, interactive jobs are limited to two hours, and long-lived AI-agent services cannot be assumed in typical shared-HPC usage. 
Although this constraint could be relaxed by external orchestration, we intentionally use this restrictive session model. 
In the controlled experiments, each run is further interrupted after 30 minutes of active work to standardize the session boundary; the agent generates a progress report, the session and interactive job are terminated, and a new session resumes from the report.

\subsection{Runtime-State Reconstruction Failures}
\label{subsec:runtime_state_failures}

The most difficult phase was the construction of CPU kernel benchmarks from runtime states. 
The \texttt{variable\_list} used in this phase is an intermediate artifact for runtime-state reconstruction. 
For each target OpenMP region, it records the variables referenced by the region, their type and array structure, their role in the region such as input or output, and any condition under which the variable is valid for dumping and replay. 
The list serves as the interface between the original simulation, dump instrumentation, and standalone benchmark generation.

We did not rely only on AI-based static analysis to enumerate the variables to be dumped. 
Instead, we intentionally used compiler errors as a low-cost discovery mechanism. 
We applied OpenMP \texttt{default(none)} to copies of the original target regions so that variables referenced inside each OpenMP region would appear in compiler diagnostics unless their data-sharing attributes were explicitly specified. 
The agent then recorded these variables, together with their original sharing attributes and access roles, in the \texttt{variable\_list}. 
This process was repeated until the copied source compiled. 
Because omissions in this variable enumeration can lead to another expensive dump run, this compiler-guided loop converts a potentially high-cost dump-time failure into low-cost compile-time feedback.

This conservative enumeration is necessary because the intermediate artifact is already large even in the reduced 15-kernel setting. 
The generated \texttt{variable\_list} files contain 690 entries, including 134 condition-dependent entries.
In the full workflow, more than 2000 dump insertion points were required, including hundreds of condition-dependent cases. 
At this scale, even a small omission rate becomes significant: a 99\% success rate would still leave roughly 20 missing or incorrect dump points.

A representative failure mode is conditional access combined with dummy allocations. 
In CReSS, when a particular physics option is disabled, a full-domain array argument may be backed by a minimal dummy object to reduce memory usage. 
The callee can still declare and use the argument through a full-domain interface, while the original code avoids invalid access through configuration-dependent control flow. 
Therefore, a local inspection of the OpenMP region may show an ordinary full-domain array access, even though dumping that array is valid only under the corresponding runtime condition. 
If dump instrumentation ignores this condition and attempts to dump the dummy-backed object as a full array, the dump run can fail with a segmentation fault.

A complete analysis of allocations, configuration branches, and caller-callee assumptions could recover such conditions more systematically, but it conflicts with the localization strategy of this workflow. 
We therefore limit reconstruction to the target routine and nearby call context, and encode required data-validity conditions in the \texttt{variable\_list}. 
This CReSS-specific pattern illustrates a broader risk: application knowledge may be hidden in option-dependent allocation or caller-callee assumptions, and may become visible only when dump instrumentation fails.
Broad refactoring could make some of them explicit, but it would also introduce behavior changes outside the target GPU kernels and require its own validation effort. 
We therefore preserve the original implementation and reconstruct only the runtime conditions needed for dump-based validation.

Such failures are difficult to diagnose in general because the observed symptom is a segmentation fault inside the dump routine, not at the source of the missing data-validity condition.
Moreover, the local OpenMP region and callee interface may look consistent because the argument is declared with a full-domain interface. 
In one run, the agent misdiagnosed the failure as a bug in the dump routine and entered a repair loop that modified the dump function rather than the \texttt{variable\_list}; the run was manually terminated after the workflow diverged. 
In less severe cases, the agent fixed one missing condition, reran the simulation, and then exposed another missing condition, leading to repeated dump re-execution.

\subsection{Specification as Externalized Workflow Context}
\label{subsec:specification_context}

To reduce such failures, we externalized workflow rules and recovery policies into specification documents. 
The specifications describe the ordering of \texttt{variable\_list} construction, dump acquisition, benchmark generation, and validation, as well as rules for handling dummy arrays, conditional accesses, benchmark configuration files, compiler options, binary-I/O conventions, and recovery policies. 
In particular, dump generation and benchmark replay must use consistent byte-order settings, such as whether \texttt{-Mbyteswapio} is enabled. 
The rules for CReSS-specific data-validity conditions, including dummy arrays and conditional accesses, were derived from failures observed during the full CReSS porting process and were included in the baseline specification for all controlled experiments; they are not an ablation target in Table~\ref{tab:prompt_vs_spec}.

The role of the specification is not to decide whether a numerical discrepancy is acceptable. 
That judgment remains with the human developer. 
Instead, the specification serves as persistent procedural context across AI-agent sessions.

This persistent context is needed because, as discussed in Section~\ref{subsec:cost_structure}, the workflow is executed through time-limited interactive jobs on a shared HPC system.

We compare two approaches for controlling Phase 3, as shown in Table~\ref{tab:prompt_vs_spec}. 
Both use the progress-report continuation protocol described in Section~\ref{subsec:cost_structure}. 
In the prompt-based setting, the full specification is given only in the initial session and later sessions resume from the generated report; in the specification-based setting, the report is used for continuation but the specification also remains available as persistent files.
The Spec column indicates whether procedural requirements such as benchmark configuration files, compiler options, byte-order settings, interactive-job execution, and recovery rules were followed.

In the prompt-based setting, convergence was achieved in three out of five runs. In one failed run, after a dump failure, the recovery rule for revisiting the \texttt{variable\_list} and checking similar data-validity conditions was not retained across session continuation; the agent misdiagnosed the failure as a dump-function bug and entered an unrecoverable repair loop. 
In another run, the agent treated partial success as sufficient and attempted to proceed before all 15 kernels were verified. 
Prompt-based runs also showed more Spec violations, where procedural requirements were weakened or lost as progress reports replaced the original prompt across sessions.

In contrast, the specification-based workflow converged in all five runs, indicating that persistent specifications help preserve procedural constraints across session boundaries. 
However, dump executions still varied from 1 to 7, which directly translates into wall-clock development cost. 
The remaining Spec violation in the specification-based runs was minor and localized, and did not affect workflow correctness. 
Thus, specifications stabilize workflow convergence and make local deviations observable, but do not by themselves provide global cost awareness.

\begin{table}[t]
\caption{Comparison of prompt-based and specification-based workflows in the reduced Phase-3 setting.}
\label{tab:prompt_vs_spec}
\centering
\footnotesize
\setlength{\tabcolsep}{3.0pt}
\renewcommand{\arraystretch}{1.05}
\begin{tabular}{llcccc}
\hline
Workflow & Run & Sessions & Dump exec. & Convergence & Spec \\
\hline
\multirow{5}{*}{Prompt-based}
 & R1 & 4$^{*}$ & -- & No  & No  \\
 & R2 & 3       & 2  & Yes & Yes \\
 & R3 & 2       & 1  & Yes & No  \\
 & R4 & 4       & 2  & Yes & No  \\
 & R5 & 4$^{*}$ & -- & No  & No  \\
\hline
\multirow{5}{*}{Spec.-based}
 & R1 & 3 & 5 & Yes & Yes \\
 & R2 & 5 & 7 & Yes & Yes \\
 & R3 & 4 & 2 & Yes & No  \\
 & R4 & 2 & 1 & Yes & Yes \\
 & R5 & 5 & 3 & Yes & Yes \\
\hline
\end{tabular}
\vspace{1mm}

\footnotesize{$^{*}$ Terminated manually due to workflow divergence or premature decision.}
\end{table}

\subsection{Cost-Aware Recovery}
\label{subsec:cost_aware_recovery}

Starting from the specification-based setting, we added a batch-checking recovery rule: after one dump failure, the agent must inspect similar variables and data-validity conditions before rerunning the simulation. 

Across five runs with this batch-error-correction specification, the session counts were 3, 2$^{*}$, 2, 2, and 1, and the corresponding dump executions were 3, --, 1, 2, and 1. 
The asterisk denotes a run terminated due to a job-control violation: despite the specification requiring interactive execution, the agent submitted the simulation as a batch job from the interactive environment. 
Because the resulting queue delay made the run unsuitable for the intended wall-clock cost measurement, we manually terminated it and excluded it from the successful-run comparison. 
Excluding this job-control failure, all successful runs completed with 1--3 dump executions, whereas the original specification-based runs in Table~\ref{tab:prompt_vs_spec} included cases with five and seven dump executions.

These results indicate that cost-aware recovery must be made explicit. 
After a high-cost failure, the agent should first search for similar failure patterns, revisit assumptions in the \texttt{variable\_list}, and avoid repeated expensive simulation jobs, rather than simply patching the local symptom and rerunning the simulation.

\subsection{Limitations of Snapshot-Based Kernel Validation}
\label{subsec:snapshot_limitations}

The kernel-level validation used in this workflow is snapshot-based. 
For each target kernel, dump data are captured at the last invocation during the 360-step validation simulation. 
This design is practical because the same dump data can be reused for CPU benchmark verification, GPU benchmark validation, and kernel-local revision.

The limitation is coverage. 
During integration, we observed a case where an AI-generated CPU benchmark omitted a branch that was not executed at the last-invocation dump point but was executed at an intermediate timestep in the full simulation. 
The snapshot-based benchmark did not expose the error; it appeared only after integration. 
We therefore returned to CPU benchmark generation, restored the omitted branch, and revalidated the corresponding CPU and GPU benchmarks.

This case shows that the current kernel artifacts validate the captured runtime states for the selected typhoon scenario, but do not certify all control paths of the 162 target kernels.
Additional scenarios or dump points can reuse these artifacts as references, but require revalidation. 
Application-level validation is therefore complementary to kernel-level validation, because it exposes errors that appear only outside the captured snapshots.

A straightforward way to improve coverage would be to dump multiple invocations or all timesteps, but this is impractical.
Even a minimal dump configuration for one test case, one MPI process, one timestep, and the target kernels produced more than 400~GB of dump data. 
Thus, scaling the workflow beyond this scenario requires a dump-data lifecycle strategy: selecting representative dump points, retaining reusable reference artifacts, and compressing or discarding raw dumps when they are no longer needed. 
The present implementation should therefore be viewed as a scenario-specific validated GPU port, not a complete GPU port of all CReSS executions.
\section{Related Work}

GPU porting of production scientific applications has been studied extensively. Directive-based approaches such as OpenACC and OpenMP target have been used to reduce accelerator porting effort while keeping the source code relatively close to the original implementation. 
GPU porting has also been reported for production-level weather and climate models such as MPAS~\cite{mpas_openacc}, WRF~\cite{wrf}, CAM-SE~\cite{camse_openacc}, NICAM~\cite{nicam_gpu}, and ICON~\cite{icon_gpu}. 
These studies demonstrate feasibility, but such efforts still require substantial manual work for kernel identification, transformation, validation, and integration. 
Our work addresses this development-process cost rather than proposing a new programming model or low-level optimization.

Recent work has explored LLMs and AI agents for HPC software development. Godoy et al.~\cite{godoy_llm_hpc} evaluated LLM-assisted code generation and auto-parallelization for fundamental HPC kernels. 
HPC-Coder~\cite{hpccoder} studied HPC-specific code completion and OpenMP pragma generation, while UniPar~\cite{unipar} evaluated LLM-based translation and compiler-feedback repair between serial, CUDA, and OpenMP programs. 
ChatHPC~\cite{chathpc_jsupercomp} targets specialized AI assistants across the HPC software stack.
These studies indicate that AI systems are becoming useful for HPC code transformation, but they primarily focus on localized transformations, limited workflow units, or assistant infrastructure.
Our workflow follows the same locality principle by reducing large-application porting to kernel-level tasks while preserving the runtime context needed for validation.

Our previous GeoFEM case study~\cite{geofem_llm} demonstrated that a CLI-based AI agent can assist GPU acceleration of a legacy Fortran application, especially for well-isolated components such as a CG solver. 
In contrast, the present work studies how AI agents can support a validation-centric GPU porting workflow for CReSS, a 250,000+ line weather simulation code, involving runtime-state reconstruction, dump-based kernel benchmarking, numerical discrepancy diagnosis, and application-level validation. 
The gap addressed in this paper is how to execute such a workflow within practical wall-clock development cost while preserving the scientific validity of the original application.
\section{Conclusion}

This study presented a validation-centric AI-assisted GPU porting workflow through a case study of CReSS, a 250,000+ line legacy weather simulation code. 
The goal was not to replace the existing model with newly generated code, but to reduce the wall-clock development time required to obtain a numerically validated GPU implementation while preserving the scientific behavior accumulated by the original application.

The workflow produced GPU implementations for 162 target kernels, satisfied application-level validation criteria, and achieved a 5.1$\times$ speedup over the 72-thread Grace CPU baseline with reasonable OpenACC performance. 
Kernel-level validation also detected five single-element discrepancies caused by floating-point and intrinsic-function differences, enabling feedback to the CReSS developers. 
These results show that AI agents can make detailed dump-based validation feasible at a scale that would be difficult to manage manually.

The case study also showed that the dominant challenges in the CReSS workflow are runtime-state reconstruction, dump acquisition, session-spanning context management, and cost-aware recovery.
While the specific failure modes observed with Claude Code Opus 4.5--4.6 may change as AI agents improve, this case suggests that similar dump-based GPU-porting workflows on shared HPC systems will still need to manage the cost of acquiring meaningful runtime states, maintaining procedural context, and avoiding repeated simulation re-execution.
Future work includes extending the validated porting coverage from the 162 kernels exercised in the current scenario to additional OpenMP regions and meteorological cases, developing scalable dump-data management strategies for continued regression validation, and selecting multiple representative dump points to improve validation coverage.
Cost-aware recovery is also important for reducing repeated simulation re-execution. More aggressive inter-kernel optimizations, such as explicit data management, asynchronous execution, kernel fusion, and communication overlap, remain important, but require validation strategies beyond the kernel-local workflow considered here.

\section*{Acknowledgments}
This work was supported by JSPS KAKENHI Grant Number JP26K14842, MEXT Feasibility Study on the Future HPCI, and the JHPCN Project under Project ID jh260061.
ChatGPT by OpenAI was used for Japanese-to-English translation and readability improvement; the authors reviewed all AI-assisted content.

\bibliographystyle{IEEEtran}
\bibliography{0_main}

\end{document}